# Structural Motif Selection in Fluorinated Metal-Organic Chalcogenides Driven by Ligand Electrostatics

Md. Saiful Islam[1], Tomoaki Sakurada,[2,3] and Yeongsu Cho[1,*]

[1]Department of Chemistry, University of Houston, Houston, TX 77204, USA
[2]Department of Materials Science and Engineering, Institute of Science Tokyo, Ookayama 2-12-9 1, Meguro-ku, Tokyo 152-8552, Japan
[3]Material Integration Laboratories, AGC Inc., Yokohama, Kanagawa 230-0045, Japan

*Corresponding author email: ycho12@uh.edu

ABSTRACT: Hybrid organic-inorganic materials enable systematic structural tuning through chemical modification of organic ligands. Predictive control, however, requires mechanistic understanding of how ligand chemistry and inorganic frameworks jointly determine structural motif selection. Metal-organic chalcogenides (MOCs), where metal-chalcogenide units are covalently bonded to organic ligands, offer an ideal platform in which ligand substitution directly alters crystal structure. Here, we investigate silver selenide-based MOCs with fluorinated phenyl ligands to elucidate governing interactions. Density functional theory with fragment-based energy analysis identifies ligand-ligand interactions as the primary energetic driver of motif selection. Symmetry-adapted perturbation theory further decomposes ligand-ligand interactions and shows that electrostatic interactions are decisive in selecting the preferred motif by selectively stabilizing specific packing arrangements. The results further show that ligand orientation controls the effectiveness of long-range electrostatic interactions, establishing a physically grounded design principle for directing structural motifs in MOCs through targeted control of ligand packing and electrostatics.

**TOC Graphic**

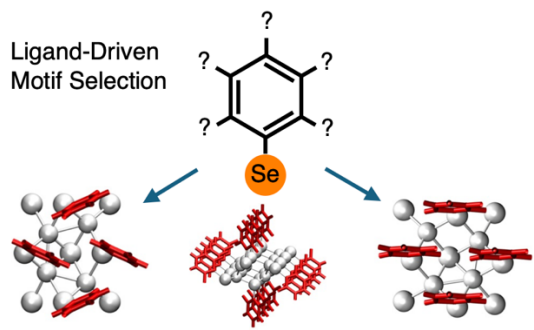

Hybrid organic-inorganic materials have attracted significant attention due to their ability to combine the robust electronic and structural properties of inorganic frameworks with the chemical tunability of organic components.[1,2] By modifying the organic components, properties such as electronic structure, interfacial energetics, and structural arrangement can be systematically tuned,[3] making these materials promising for a wide range of applications, including optoelectronics,[4,5] catalysis,[6,7] and energy conversion.[8,9] However, despite this tunability, rational design of the organic component remains challenging.[10] Small changes in molecular structure can lead to substantial differences in packing behavior and overall crystal structure, and predictive guidelines that link molecular properties to structural outcomes are still limited.[11,12]

Metal-organic chalcogenides (MOCs) have recently emerged as a versatile class of hybrid organic-inorganic materials in which metal-chalcogen frameworks are covalently bonded to organic ligands.[13,14] The strong coupling between the inorganic framework and the organic ligands makes MOCs particularly sensitive to ligand modifications.[15] As a result, variations in ligand structure can induce significant changes in the overall crystal structure, including transformations in dimensionality from zero-dimensional clusters[16] to one-[17] and two-dimensional[18] extended networks. This sensitivity makes MOCs an ideal platform for probing how molecular-level modifications influence structural motifs and for developing design principles that connect ligand properties to crystal structure.[19]

The relationship between intermolecular interactions and crystal structure has been extensively studied in molecular crystals, where noncovalent interactions determine packing arrangements.[20] While similar principles are expected to apply to hybrid organic-inorganic materials,[21] additional complexity arises because both inorganic frameworks and organic components contribute to structural determination. Elucidating how organic and inorganic components jointly determine crystal structure, together with identification of the physical origins of the governing interactions, remains critical for developing predictive design strategies.

In this work, the relationship between noncovalent interactions and structural motif selection in MOCs is systematically investigated. To isolate the role of electronic effects and minimize steric contributions, the study focuses on a series of silver selenide MOCs with fluorinated phenyl ligands, where fluorine substitution introduces minimal steric perturbation. Density functional theory (DFT) calculations are used to evaluate the contributions of different structural components. Symmetry-adapted perturbation theory (SAPT) is employed to decompose intermolecular interactions into dispersion, electrostatic, induction, and exchange contributions, enabling direct identification of the physical origin that govern structural motif selection.[22]

We study six MOCs which have been synthesized and structurally characterized by single-crystal X-ray diffraction (scXRD). These compounds include the unsubstituted system Ph[23], the mono-fluorinated derivative F(3)[24], and four difluorinated derivatives $F_2(2,3)$, $F_2(2,4)$, $F_2(2,5)$[25], and $F_2(2,6)$[26]. Based on the dimensionality of the AgSe framework and the packing arrangement of the organic ligands, the crystal structures of these compounds can be classified into three structural motifs: 2D-HB, 2D-P, and 1D-chain (Table 1). While compounds sharing the same structural motif can crystallize in different space groups, the analysis focuses on how ligand substitution governs local coordination and packing features captured by the motif classification, rather than global crystallographic symmetry.

**Table 1.** Space group and structural motif classifications of the studied MOCs.

| Label | Space group | Structural motif |
| --- | --- | --- |
| Ph[23] | $P2_1/c$ | 2D-HB |
| F(3)[24] | $P2_12_12_1$ | 2D-HB |
| $F_2(2,3)$[25] | $P2_1/n$ | 2D-P |
| $F_2(2,4)$[25] | $P\bar{1}$ | 2D-P |
| $F_2(2,5)$[25] | $P\bar{1}$ | 2D-P |
| $F_2(2,6)$[26] | $P2_1/n$ | 1D-chain |

In the 2D-HB motif, adopted by Ph and F(3), the AgSe framework forms a two-dimensional layer, and the organic ligands exhibit a herringbone packing arrangement (Figures 1a and S1a). Within the AgSe layer, each Ag atom is coordinated by four Se atoms and has three nearest Ag neighbors with similar Ag-Ag distances, resulting in a relatively uniform Ag-Ag connectivity. In contrast, $F_2(2,3)$, $F_2(2,4)$, and $F_2(2,5)$ adopt the 2D-P motif, in which the AgSe framework also forms a two-dimensional layer, while the organic ligands are arranged parallel along a single direction (Figures 1b, S1b, and S1c). Each Ag atom remains coordinated by four Se atoms and has four neighboring Ag atoms. Two of Ag neighbors are approximately 1 Å farther than the other two, producing a distorted Ag-Ag network within the layer.[25] A distinct structure is observed for $F_2(2,6)$, which adopts the 1D-chain motif (Figure 1c). In this motif, the AgSe framework forms one-dimensional chains. Each Ag atom is coordinated by three Se atoms, forming a chain-like AgSe network. The organic ligands surround the AgSe chains, and ligands belonging to neighboring chains are arranged parallel to one another.

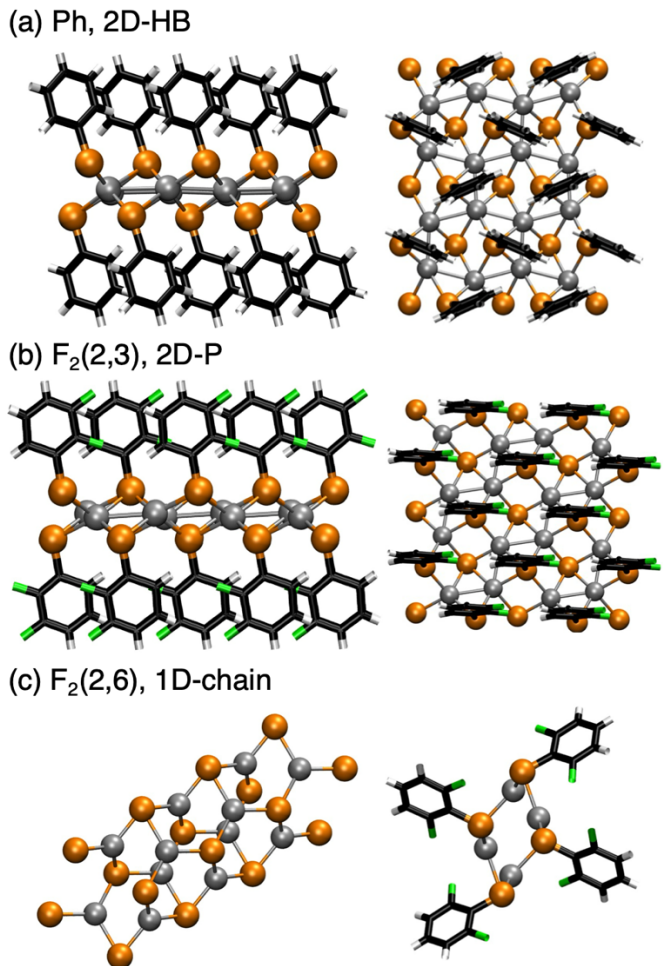

**Figure 1**. Side view (left) and top view (right) of (a) Ph, which adopts the 2D-HB motif, and (b) $F_2(2,3)$, which adopts the 2D-P motif. (c) Side view (left) of the 1D AgSe chain in $F_2(2,6)$ and front view (right) of $F_2(2,6)$, which adopts 1D-chain motif. Atoms are colored as follows: Ag, gray; Se, orange; C, black; H, white; F, green.

To evaluate how the structural motif influences the stability of each compound, a set of hypothetical structures was constructed in which a ligand A adopts the geometry observed for ligand B, denoted A[B]. The notation A[A] therefore represents the experimentally observed structure determined by scXRD. Although multiple ligands may share the same structural motif, differences in substitution lead to variations in detailed geometry within a given motif, which can influence the relative stability of the structures. All experimentally observed and hypothetical structures were fully optimized including the cell parameters while maintaining the corresponding structural motif. The relative stability of each structure was evaluated using the energy difference $\Delta E(A[B]) = E(A[B]) - E(A[A])$ (Figure 2a). For all ligands, the experimentally observed motif corresponds to the lowest-energy structure among the tested motifs. Although several hypothetical structures of $F_2(2,5)$ are lower in energy than the experimentally observed structure, they all retain the same structural motif, 2D-P. These results demonstrate that the experimentally observed motifs generally correspond to energetically preferred motifs of the MOCs.

To identify the physical origin of these stability differences, the interaction energies were partitioned into three components: AgSe-AgSe interactions within the AgSe framework, ligand-ligand interactions among organic ligands, and AgSe-ligand interactions between the AgSe framework and ligands. Contributions from AgSe-AgSe and ligand-ligand interactions were estimated by separating the MOC structures into AgSe-only and ligand-only complexes.

The AgSe-only complexes were constructed by replacing the organic ligands with hydrogen atoms and optimizing only the hydrogen positions while keeping the AgSe framework fixed (Figure S2a). The contribution of AgSe-AgSe interactions to the stability of the full MOC structures was evaluated by comparing the total energies of the AgSe-only complexes derived from the experimentally observed structures for different ligands. The resulting energies show relatively small variation across motifs, with a total range within 0.012 eV per formula unit (Figure 2b). Despite this small variation, systematic differences can still be identified. The 2D-HB motif, characterized by the most uniform Ag-Ag connectivity, yields the lowest AgSe framework energy. In contrast, the 1D-chain motif, where each Ag atom is coordinated by three Se atoms rather than four as in the two-dimensional motifs, exhibits the highest AgSe framework energy.

The ligand-only complexes were constructed by removing the AgSe framework and terminating the ligand with hydrogen atoms placed at the Se positions, followed by optimization of the hydrogen atoms (Figures S2b). The ligand-ligand interaction energy was evaluated by comparing the energy of the ligand within the ligand-only complex derived from the experimentally observed structures to that of the corresponding isolated ligand molecule (Figure 2c). Among the studied systems, Ph exhibits the strongest ligand-ligand interaction energy. Although F(3) adopts the same 2D-HB motif and has nonzero molecular dipole moment, its ligand-ligand interactions are weaker. The difluorinated ligands show more diverse behavior, with $F_2(2,3)$ and $F_2(2,5)$ showing relatively strong ligand-ligand interactions, whereas $F_2(2,4)$ and $F_2(2,6)$ show weaker interactions.

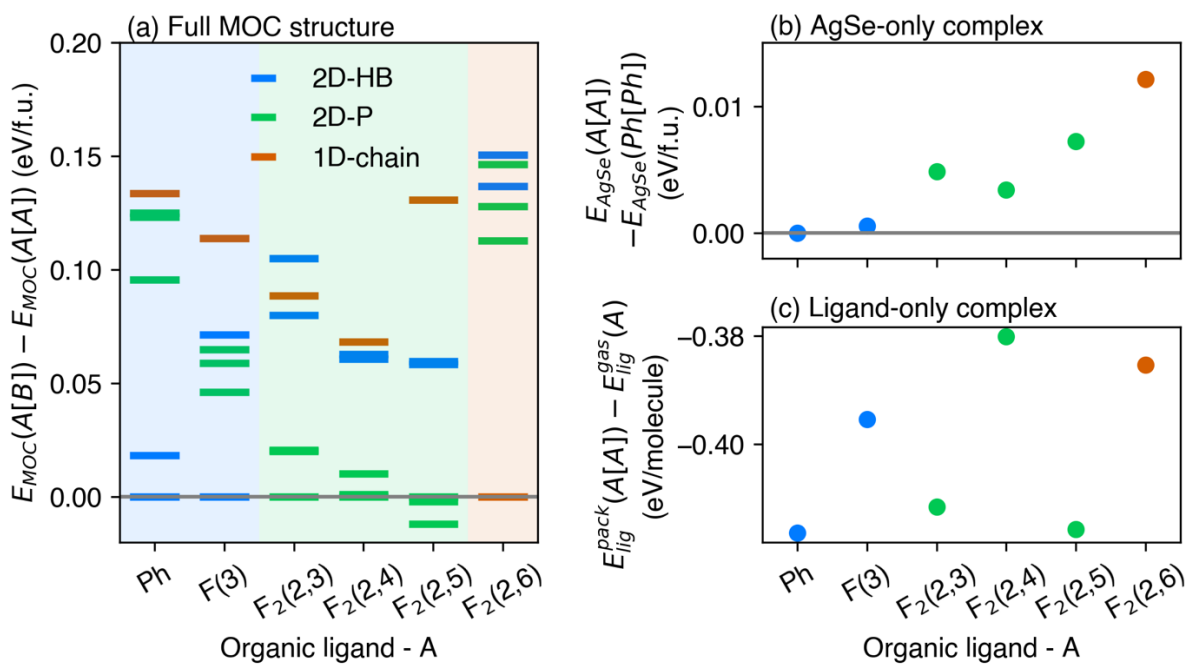

**Figure 2**. (a) Relative energies per formula unit, $E_{MOC}(A[B]) - E_{MOC}(A[A])$, of hypothetical structures A[B] referenced to the experimentally observed structure A[A], which is set to 0 eV. A[B] denotes a structure in which ligand A adopts the geometry observed for ligand B, while A[A] corresponds to the experimentally determined structure. Background shading indicates the experimentally observed motif of ligand A, and line colors indicate the motif imposed in the hypothetical structure. Positive values indicate structures less stable than the experimental motif, whereas negative values indicate more stable structures. (b) Relative energies per formula unit of the AgSe-only complexes referenced to the unsubstituted system Ph, which is set to 0 eV. $E_{AgSe}(A[A])$ denotes the total energy of the AgSe-only complex derived from the MOC containing ligand A. (c) Ligand-ligand interaction energies obtained from the difference between the energy of a ligand in the ligand-only complex, $E_{lig}^{pack}$, and that of the isolated ligand, $E_{lig}^{gas}$.

To further examine the role of each interaction component, the contribution of AgSe-AgSe and ligand-ligand interaction energies to the relative stability of the full MOC structures were evaluated using AgSe-only and ligand-only complexes derived from hypothetical structures adopting different structural motifs, with energies referenced from those derived from the experimentally observed structures. The variation in AgSe-AgSe interaction energy across motifs remains small, less than 0.03 eV per formula unit (Figure 3a), whereas the energy differences among hypothetical structures of the full MOC structures reach 0.15 eV per formula unit (Figure 2a). This comparison indicates that variations in AgSe framework energy alone cannot account for the observed motif preferences. In contrast, ligand-ligand interaction energies exhibit substantially larger variations across motifs, also reaching 0.15 eV (Figures 3b). Moreover, the trend in ligand-ligand interaction energies closely follows the relative stability of the full MOC structures in that the lowest-energy motif corresponds to the experimentally observed motif (Figure 2a), indicating that ligand packing plays a major role in determining the preferred structural motif.

Although AgSe-ligand interactions cannot be isolated in the same manner as the AgSe-AgSe and ligand-ligand contributions, an effective AgSe-ligand interaction energies can be estimated by subtracting the relative energies of the AgSe-only and ligand-only complexes from the relative energy of the full MOC structures. The resulting variation in the effective AgSe-ligand interaction remains small, less than 0.07 eV across motifs, and does not correlate with the lowest-energy motif of the full MOC structures (Figure 3c), indicating that AgSe-ligand interactions do not play a decisive role in motif selection. Overall, the preferred structural motif is primarily determined by ligand-ligand interactions that favor specific packing arrangements, while the AgSe framework imposes geometric constraints that stabilize arrangements that best accommodate these ligand-ligand interactions.

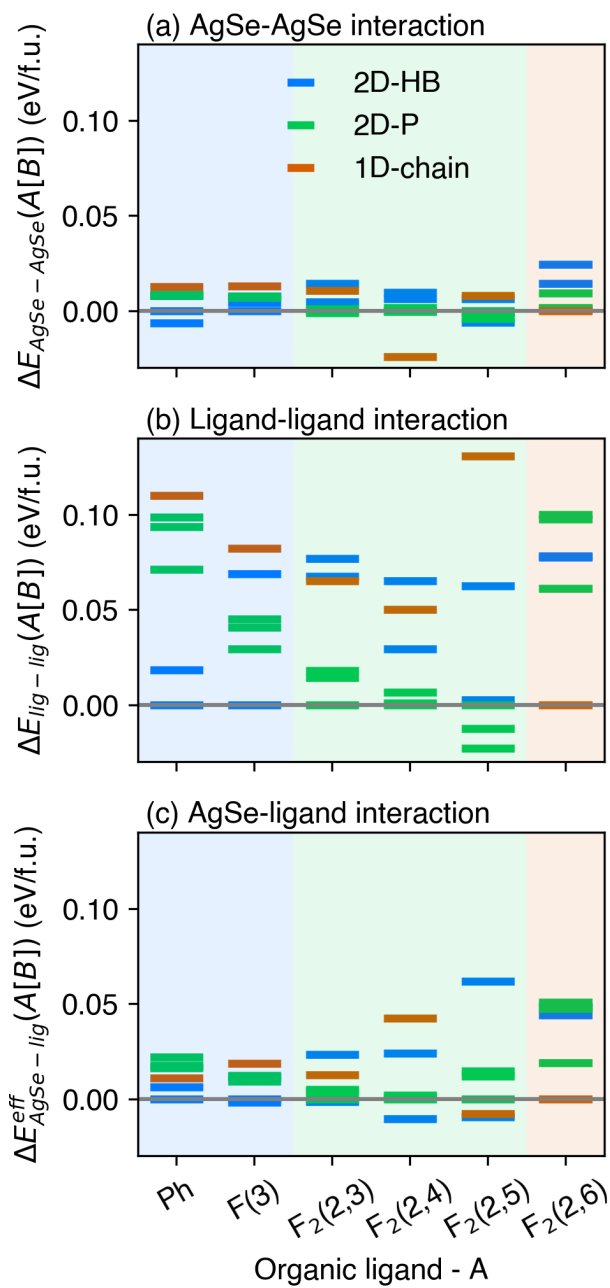

**Figure 3**. Relative energies per formula unit of (a) AgSe-only complex and (b) ligand-only complex of hypothetical structures A[B] referenced to the experimentally observed structure A[A], which is set to 0 eV. A[B] denotes a structure in which ligand A adopts the geometry observed for ligand B, while A[A] corresponds to the experimentally determined structure. (c) Effective relative energies of AgSe-ligand interactions, estimated as $\Delta E_{MOC} - \Delta E_{AgSe-AgSe} - \Delta E_{lig-lig}$, where $\Delta E_{MOC}$, $\Delta E_{AgSe-AgSe}$, and $\Delta E_{lig-lig}$ correspond to the relative energies of the full MOC, AgSe-

only, and ligand-only complexes, respectively. Background shading indicates the experimentally observed motif of ligand A, and line colors indicate the motif imposed in the hypothetical structure.

Building on the observation that ligand-ligand interactions govern motif selection, SAPT was applied to decompose the intermolecular interactions between ligands into physically meaningful contributions. The SAPT calculations were performed on the ligand-only complexes to partition the interaction energy into dispersion, electrostatic, induction, and exchange components. For each structure, SAPT interaction energies were evaluated for all ligand pairs separated by less than 10 Å, and the total contribution for ligand-ligand interactions of a given structure was obtained by summing the pairwise contributions.

Three considerations are important when interpreting the pairwise SAPT analysis. First, the electronic environment of the ligand differs between the full MOC and the ligand-only complex due to removal of the Se-C covalent bond. Analysis on the sixth-generation density derived electrostatic and chemical (DDEC6) partial charges[30–32] indicates that this modification has a localized effect only, with most atomic charges remaining largely unchanged, while the C atom previously bonded to Se becomes more negative by 0.18 on average (Figure S3). Second, SAPT provides a pairwise decomposition of ligand-ligand interactions and therefore does not fully include depolarization effects arising from the electrostatic environment imposed by ligand packing in the full MOC. Third, quantitative differences relative to periodic DFT are expected because SAPT and periodic DFT employ distinct electronic structure methods and basis representations.

The reliability of the pairwise SAPT analysis was therefore assessed by comparing the total interaction energies obtained from SAPT with those calculated using periodic DFT. A comparison of ligand-ligand interaction energies for all experimental and hypothetical structures shows that the SAPT interaction energies are systematically more attractive than the periodic DFT energies by 0.15 eV on average, while preserving the overall energetic trends (Figure S4). Furthermore, the relative total interaction energies obtained from SAPT (Figure S5) follow trends similar to those obtained from periodic DFT (Figures 2a and 3b), with the lowest energy motif in each case corresponding to the experimentally observed motif. Preservation of energetic trends indicates that the dominant interaction components governing motif selection are properly captured by SAPT despite the charge redistribution associated with removal of the Se-C bond, the absence of explicit depolarization, and differences in methodology. The influence of depolarization arising from ligand packing is further discussed below.

Among the SAPT components, dispersion provides the dominant attractive contribution to ligand-ligand interactions, accounting for 73% of the total attractive interaction energy on average (Figure S6). The 2D-HB motif generally exhibits stronger dispersion interactions than the alternative motifs across the different ligands (Figures 4a). The herringbone arrangements allow neighboring aromatic rings to approach each other with favorable orientations that enhance intermolecular $\pi$-$\pi$ dispersion interactions.[27] Consistent with this picture, for Ph and F(3), the experimentally observed 2D-HB motif yields the strongest dispersion stabilization. For $F_2(2,3)$, $F_2(2,4)$, and $F_2(2,5)$, the dispersion energies of the 2D-HB and 2D-P motifs are comparable, although 2D-HB remains slightly more favorable for $F_2(2,4)$ and $F_2(2,5)$. Even for $F_2(2,6)$, the 2D-HB motif exhibits stronger dispersion stabilization than the other motifs, while the 1D-chain motif becomes more favorable than the 2D-P motif. These results indicate that dispersion interactions primarily determine the overall magnitude of ligand-ligand stabilization but do not uniquely select the preferred structural motif.

Electrostatic interactions also contribute substantially to motif stabilization, accounting for 20% of the attractive interaction energy on average (Figure S6). Although smaller in magnitude than the dispersion contribution, electrostatic interactions consistently favor the experimentally observed motif for each ligand (Figures 4b). In particular, for $F_2(2,6)$, the 1D-chain exhibits substantially stronger stabilization exceeding 0.25 eV relative to the two-dimensional motifs. This behavior reflects the directional nature of electrostatic interactions, which preferentially stabilize specific relative orientations of adjacent ligands and thereby play a decisive role in selecting the preferred structural motif.

In contrast, induction interactions contribute only weakly to the total interaction energy, representing 7% of the attractive interactions on average (Figures S6). The induction contribution varies little across different motifs (Figure 4c), reflecting the absence of strong donor-acceptor interactions between neighboring ligands. Consequently, induction interactions play only a minor role in determining the preferred structural motif.

Exchange interactions show the opposite trend relative to the attractive interactions (Figure 4d). In all cases, the experimentally observed motifs exhibit less favorable exchange energies than the alternative motifs. This is because maximizing attractive interactions typically requires ligands to approach each other closely, which inevitably increases orbital overlap and therefore exchange repulsion. The preferred structural motifs reflect a balance between maximizing attractive interactions while tolerating the associated increase in exchange repulsion. Overall, the SAPT analysis shows that dispersion interactions dominate the stabilization, whereas electrostatic interactions determine the preferred ligand orientation, leading to the experimentally observed structural motifs. This conclusion is consistent with previous studies on aromatic dimers, where dispersion provides the primary stabilization while electrostatic interactions determine the preferred relative orientation.[28,29]

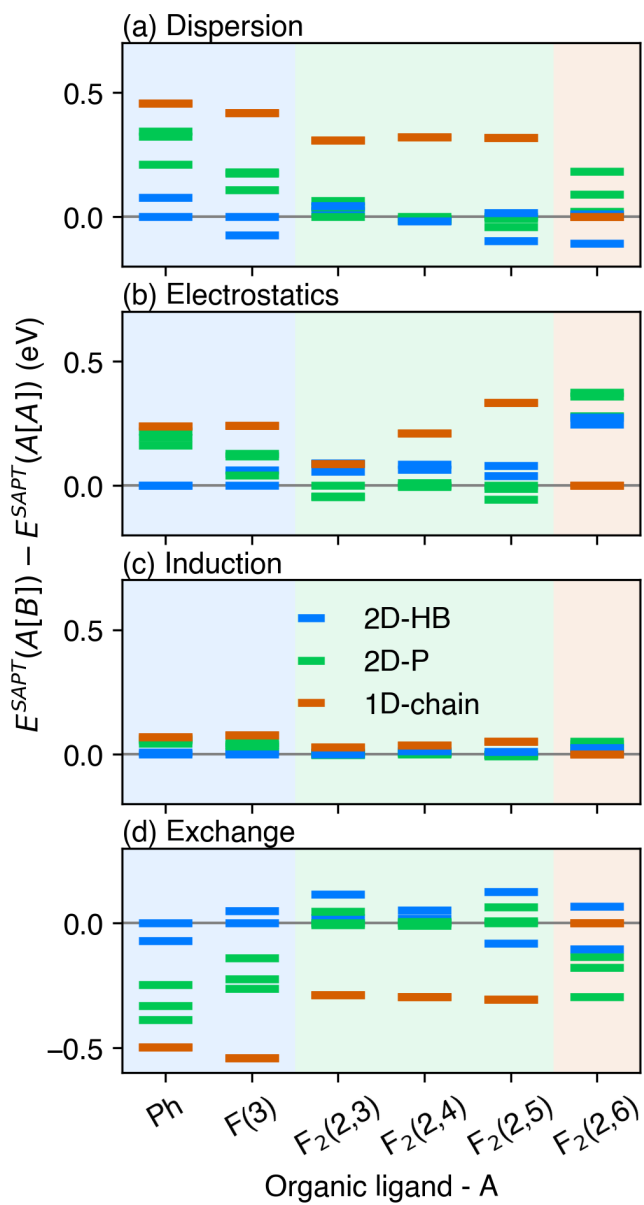

**Figure 4**. Relative SAPT interaction energy components (a) dispersion, (b) electrostatics, (c) induction, and (d) exchange for hypothetical structures A[B] with respect to the experimental structure A[A], which is set to 0 eV. Background shading indicates the experimentally observed motif of ligand A, and line colors indicate the motif imposed in the hypothetical structure. Positive values indicate weaker interactions than in the experimental motif, whereas negative values indicate stronger interactions.

The 1D-chain motif exhibits distinct behavior compared to the two-dimensional motifs, motivating a more detailed analysis of its ligand-ligand interactions. For most ligands, the 1D-chain motif shows notably weaker dispersion and electrostatic interactions, making it energetically unfavorable. The only exception is the $F_2(2,6)$ ligand, which experimentally adopts the 1D-chain motif. To understand this behavior, the SAPT interaction energies of individual ligand pairs were analyzed as a function of the distance between ligand centers for structures with the 1D-chain motif (Figure S6). Most ligand pairs exhibit a distance dependence consistent with dispersion-dominated interactions, approximately scaling as $1/r^6$. However, two pairs of $F_2(2,6)$ and three pairs of $F_2(2,3)$ significantly deviate from this trend and exhibit unusually strong long-range interactions. In the case of $F_2(2,6)$, the two ligands are arranged nearly coplanar with opposite orientations, which enables effective long-range electrostatic interactions between the dipole moments of the two ligands. For $F_2(2,3)$, the molecular dipole moment is larger, and the dipoles are aligned in a configuration that also allows favorable electrostatic interactions at intermediate separations. However, the ligand pairs with the shortest distances in $F_2(2,3)$ exhibit relatively weak interactions, preventing the overall ligand-ligand interactions from favoring the 1D-chain motif. Consequently, only $F_2(2,6)$ adopts the 1D-chain motif experimentally. These results indicate that fluorine substitution not only modulates intermolecular interactions through changes in dipole moment or polarizability but also determines how effective these interactions can be realized within a given structural motif due to orientations of the substituents.

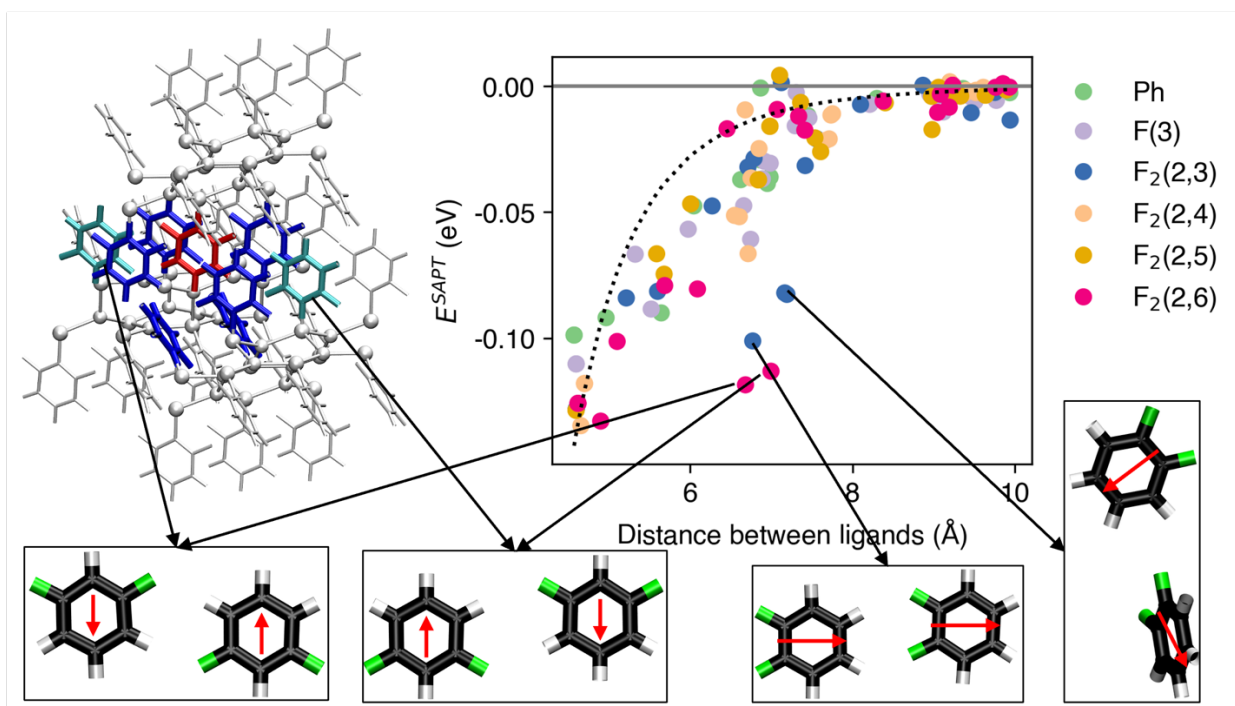

**Figure 5**. Analysis of ligand-ligand interactions in the 1D-chain motif. The left panel shows eight ligand pairs with the shortest intermolecular distances in $F_2(2,6)$. The origin of the pair is indicated in red, and the six closest pairs are indicated in blue. Two pairs exhibiting unusually strong long-range interactions are indicated in cyan. The remaining MOC framework is shown in white bonds, and Ag and Se atoms are represented by white spheres. The right panel shows total SAPT interaction energies as a function of the distance between ligand centers for ligand pairs separated by less than 10 Å in the 1D-chain motif. Each point corresponds to a ligand pair, and colors indicate the ligand composition. Black dotted line represents a $1/r^6$ fit. Structures corresponding to the two pairs of $F_2(2,6)$ and three pairs of $F_2(2,3)$ that exhibit strong long-range interactions are illustrated below. The rightmost structure illustrates two symmetry-equivalent pairs of $F_2(2,3)$. Molecular dipole moments are indicated by red arrows.

Although SAPT provides valuable insights into the pairwise components of ligand-ligand interactions, it does not fully capture collective polarization effects that arise in the packed crystal. The dipole moments of neighboring molecules can induce redistribution of charge density within each ligand, leading to depolarization. To examine the extent of this effect, dipole moments were calculated for isolated ligands and for ligands in the ligand-only complexes using DDEC6 partial charges, providing an approximate trend (Figure S7). Both Ph and $F_2(2,5)$ have zero dipole moment in the isolated state, but nonzero dipole moments are induced upon packing. The induced dipole is larger for $F_2(2,5)$, reflecting its greater molecular polarizability. For ligands that already possess a permanent dipole moment, $F(3)$, $F_2(2,3)$, $F_2(2, 4)$, and $F_2(2,6)$, packing generally reduces the dipole magnitude by 0.07-0.74 D due to depolarization. This effect is particularly pronounced for $F_2(2,6)$, whose dipole decreases by more than 0.69 D in all two-dimensional motifs but remains enhanced in the 1D-chain motif. The reduced depolarization therefore provides an additional electrostatic stabilization that favors formation of the 1D-chain motif for $F_2(2,6)$. Overall, depolarization complements the SAPT analysis by showing that, although not always dominant, collective polarization effects associated with ligand packing orientation can further influence motif selection.

In summary, the relationship between structural motifs and noncovalent interactions in silver selenide-based MOCs with fluorinated phenyl ligands was systematically investigated for six compounds, Ph, $F(3)$, $F_2(2,3)$, $F_2(2,4)$, $F_2(2,5)$, and $F_2(2,6)$. By comparing the relative contributions of AgSe-AgSe, AgSe-ligand, and ligand-ligand interactions using the AgSe-only and ligand-only complexes, ligand-ligand interactions were identified as the dominant factor governing motif selection, whereas the AgSe framework contributes only minor energetic variation across motifs. Structural motifs are therefore selected by ligand-ligand interactions that drive packing preferences, with the AgSe framework acting primarily as a geometrical scaffold that constrains the accessible geometries over which ligand-ligand stabilization is optimized.

A detailed decomposition of ligand-ligand interactions using SAPT revealed that, although dispersion interactions contribute the majority of the stabilization, electrostatic interactions play the decisive role in determining the preferred structural motif. The directional nature of electrostatic interactions enables selective stabilization of specific packing arrangements, leading to the experimentally observed motifs.

Analysis of fluorinated ligands further demonstrates that the magnitude of the molecular dipole moment alone does not fully determine the packing behavior. The relative orientation of neighboring ligands critically influences whether dipole-dipole interactions enhance or reduce stabilization. In particular, $F_2(2,6)$ adopts the 1D-chain motif because its molecular arrangement enables effective dipole-dipole interactions while minimizing depolarization, thereby enhancing electrostatic stabilization of 1D-chain motif relative to competing motifs.

These findings establish a physically grounded design principle in which both molecular dipole characteristics and packing geometry must be considered simultaneously. Future ligand design can therefore move beyond simple substituent-based tuning of dipole moments toward strategies that explicitly encode favorable intermolecular orientations and collective polarization effects. Extending this framework to systems with larger or more complex substituents will provide opportunities to exploit the interplay between steric and electrostatic interactions to access new structural motifs and properties. More broadly, the insights developed here offer a general strategy for controlling structure in hybrid organic-inorganic materials through targeted manipulation of intermolecular interactions. Such an approach has the potential to enable predictive design of hybrid organic-inorganic materials, where precise control over molecular packing is essential for tuning electronic, optical, and interfacial properties.

**Computational Details**

All DFT calculations were carried out using the Vienna Ab initio Simulation Package (VASP) version 6.5.1.[33–36] Brillouin zone sampling was determined using a k-point density of 4000 per number of atoms per unit cell, as determined by the k point density generator in Pymatgen.[37] The Perdew-Burke-Ernzerhof (PBE) functional[38] was used in conjunction with projector-augmented wave (PAW) pseudopotentials,[39] and a plane-wave kinetic energy cutoff of 800 eV.

Dispersion interactions were accounted for using Grimme's DFT-D3 correction[40] with Becke-Johnson damping[41]. Electronic self-consistent field iterations were converged to $10^{-4}$ eV, and structural relaxation proceeded until the total energy change between ionic steps fell below $10^{-3}$ eV. Energies of isolated molecules were calculated in a 30 Å × 30 Å × 30 Å simulation cell to eliminate spurious periodic interactions. Partial atomic charges for the dipole moment calculations were obtained using the sixth-generation density derived electrostatic and chemical (DDEC6) partitioning scheme, based on the DFT charge density and a code published on Source Forge.[30–32] Symmetry-adapted perturbation theory (SAPT) calculations[22] were performed at the SAPT2+3 level of theory[42] using Psi4 version 1.9.1[43] with the aug-cc-pVTZ basis set.[44,45]

## ASSOCIATED CONTENT

**Supporting Information.**

The Supporting Information is available free of charge at https://pubs.acs.org.

Structure of F(3), $F_2$(2,4), and $F_2$(2,5); Structure of AgSe-only and ligand-only complex of Ph; Partial charges of ligand atoms in full MOCs and ligand-only complexes; Ligand-ligand interaction energies calculated using periodic DFT and SAPT; Relative ligand-ligand interaction energies obtained from SAPT; Four components of SAPT interaction energies; Dipole moments of ligand in isolated and packed states. (PDF)

## AUTHOR INFORMATION

**Corresponding Author**

*email: ycho12@uh.edu
**Corresponding Author**

*email: ycho12@uh.edu

**Notes**

The authors declare no competing financial interest.

ACKNOWLEDGMENTS

The authors acknowledge the Texas University Fund for supporting start-up funding at the University of Houston. This work was completed with resources provided by the Research Computing Data Core at the University of Houston.

**Supporting Information for**

*Structure Motif Selection in Fluorinated Metal-Organic Chalcogenides Driven by Ligand Electrostatics*

Md. Saiful Islam[1], Tomoaki Sakurada,[2,3] and Yeongsu Cho[1,*]

[1]*Department of Chemistry, University of Houston, Houston, TX 77204, USA*
[2]*Department of Materials Science and Engineering, Institute of Science Tokyo, Ookayama 2-12-9 1, Meguro-ku, Tokyo 152-8552, Japan*
[3]*Material Integration Laboratories, AGC Inc., Yokohama, Kanagawa 230-0045, Japan*

**Contents**

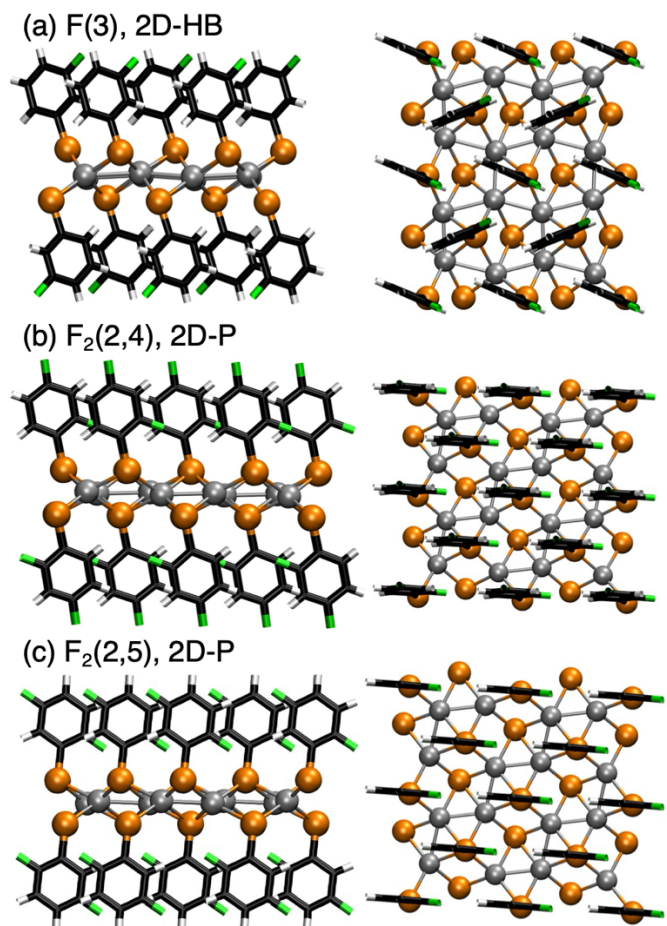

**Figure S1**. Side view (left) and top view (right) of (a) F(3), which adopts the 2D-HB motif, and (b) $F_2(2,4)$ and (c) $F_2(2,5)$, which adopts the 2D-P motif. Atoms are colored as follows: Ag, gray; Se, orange; C, black; H, white; F, green.

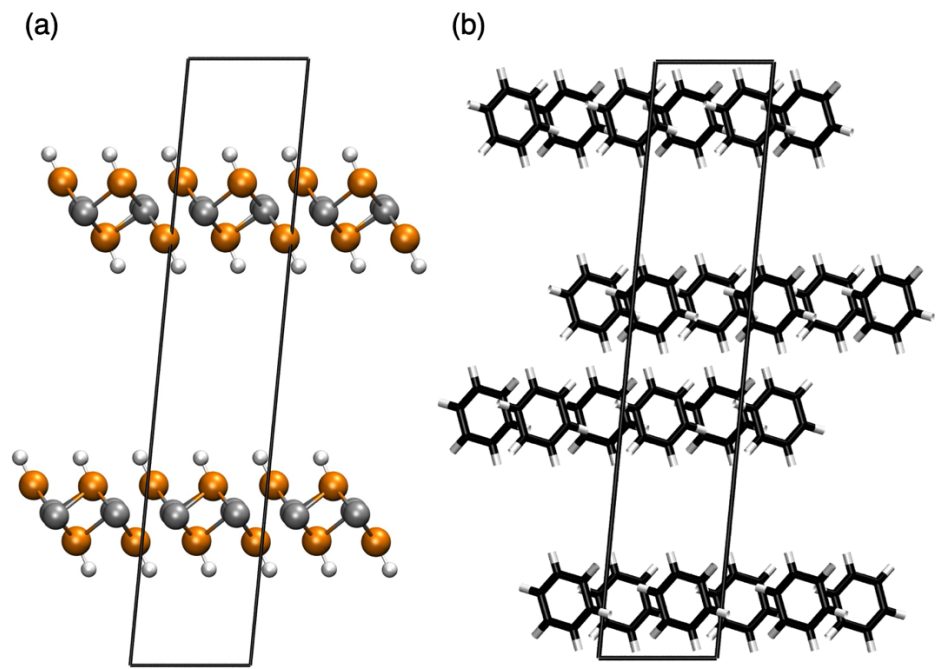

**Figure S2**. (a) AgSe-only complex and (b) ligand-only complex of Ph. Black line indicates the unit cell. Atoms are colored as follows: Ag, gray; Se, orange; C, black; H, white.

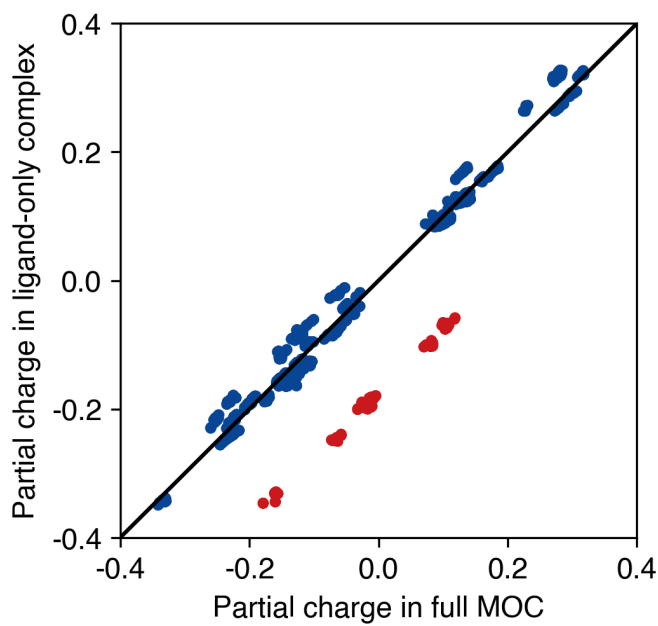

**Figure S3**. The sixth-generation density derived electrostatic and chemical (DDEC6) partial charges of ligand atoms in full MOCs and corresponding ligand-only complexes for experimentally observed and hypothetical structures. Red points denote carbon atoms bonded to Se in the full MOC. Black solid line represents equality of partial charge between the two environments.

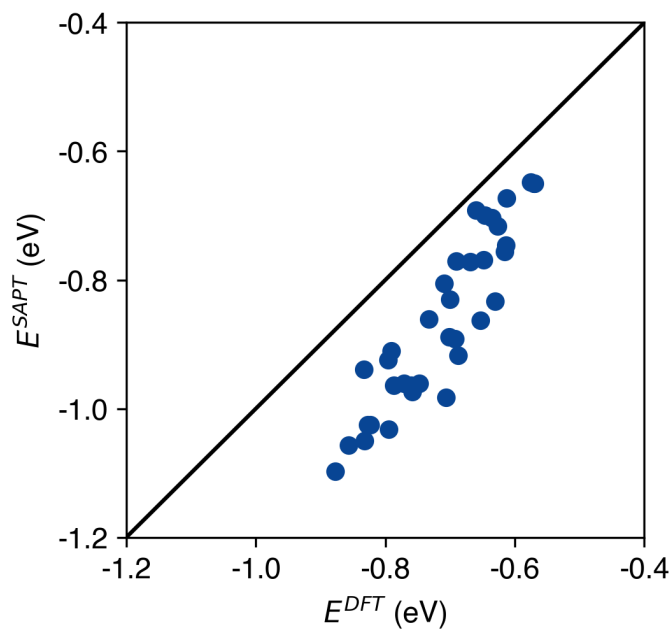

**Figure S4**. Comparison of ligand-ligand interaction energies obtained from periodic DFT and SAPT calculations for experimentally observed and hypothetical structures. $E^{DFT}$ is calculated as the energy of one ligand in the ligand-only complex minus the energy of the corresponding isolated ligand, multiplied by two to represent the interaction energy of a ligand pair. $E^{SAPT}$ obtained by summing SAPT interaction energies for all ligand pairs separated by less than 10 Å in the ligand-only complex. Black solid line indicates $E^{DFT} = E^{SAPT}$.

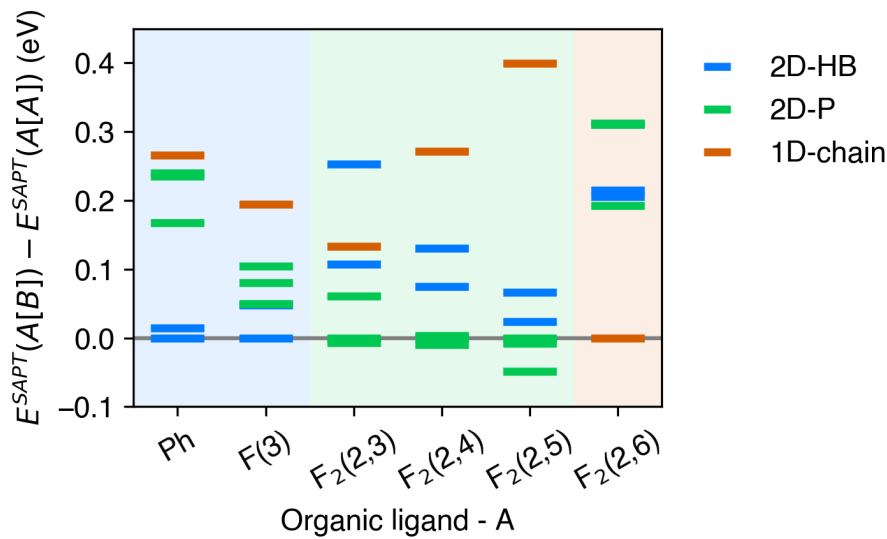

**Figure S5**. Relative total ligand-ligand interaction energies obtained from SAPT for ligand-only complexes. The total interaction energy, $E^{SAPT}$, is obtained by summing SAPT interaction energies for all ligand pairs separated by less than 10 Å in the ligand-only complex. Energies of hypothetical structures ($E^{SAPT}(A[B])$) are shown relative to the experimentally observed structures ($E^{SAPT}(A[A])$), which is set to 0 eV. A[B] denotes a structure in which ligand A adopts the structural motif observed for ligand B, while A[A] corresponds to the experimentally determined structure. Background shading indicates the experimentally observed motif of ligand A, and line colors indicate the motif imposed in the hypothetical structure.

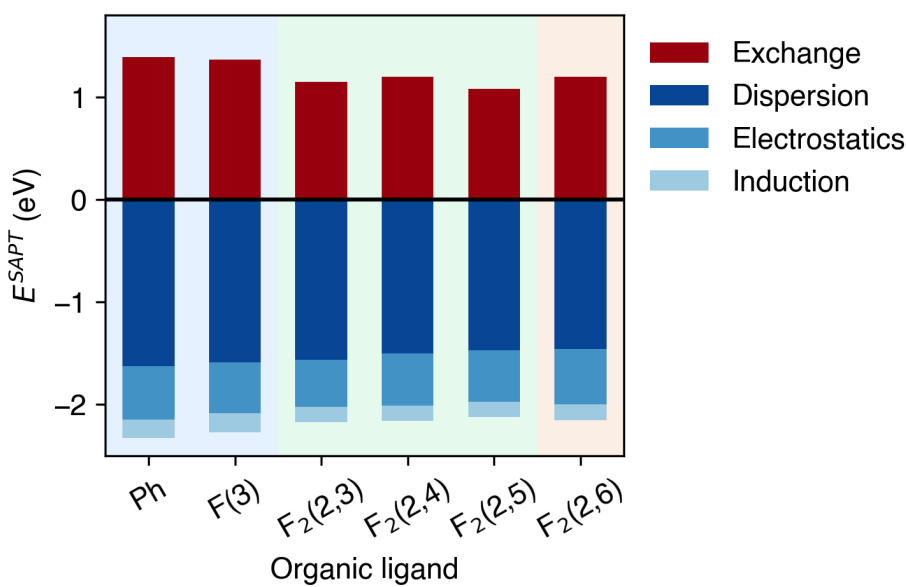

**Figure S6**. Four components of SAPT interaction energies of experimentally observed structures. Background shading indicates the motif of the ligand.

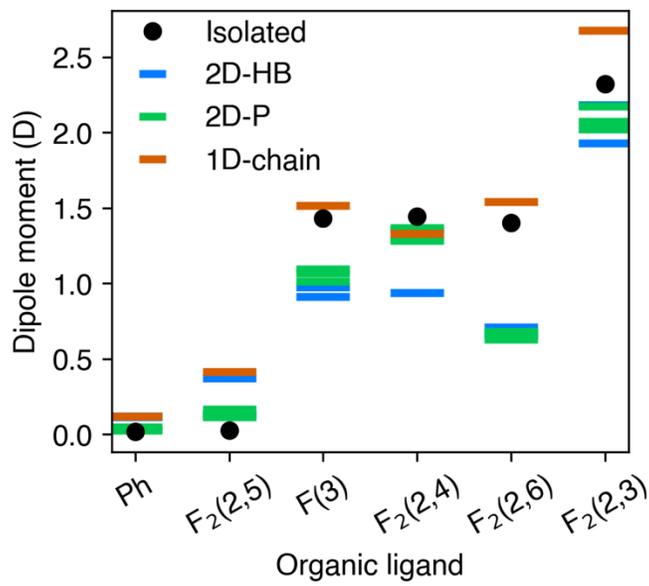

**Figure S7**. Molecular dipole moments of organic ligands calculated using DDEC6 partial charges. Black circles indicate dipole moments of isolated ligands, while horizontal lines represent dipole moments of ligands in packed ligand-only complexes, colored according to the structural motif.